# The First Estimates of Kinematically Forbidden *D* Meson Decays

R.C. Verma[1,2], Norikazu Yamada[1,3] and Kosuke Odagiri [4]

[1]KEK Theory Center, High Energy Accelerator Research Organization (KEK), Tsukuba 305-0801, Japan

[2]Panjab University, Chandigarh, India -160 014

[3]Graduate University for Advanced Studies (SOKENDAI), Tsukuba 305-0801, Japan

[4]Electronics and Photonics Research Institute, National Institute of Advanced Industrial Science and Technology, Tsukuba Central 2, 1–1–1 Umezono, Tsukuba, Ibaraki 305–8568, Japan

**Abstract**

The weak hadronic decay $D^+ \to \bar{K}^{*0} a_1^+$ is kinematically forbidden at the peak mass values of the particles involved. However, occurrence of this decay has been reported with branching fraction $(9.1 \pm 1.8) \times 10^{-3}$ in the analysis of $D^+ \to \bar{K} \, 4\pi$ decay data. This is due to smearing effects on this decay caused mainly by the large width of $a_1$-resonance, which extends the phase space and allows this decay. Using a factorization model to evaluate decay amplitudes for external and internal W-emission diagrams, and incorporating Breit-Wigner smearing using the total $a_1$ width of 400 MeV, we obtain the first estimate for branching fraction of this decay to be $3.3 \times 10^{-3}$ and $7.0 \times 10^{-3}$, for $|V_1^{Da_1}(0)| = 0.40$ and $1.50$ respectively corresponding to different theoretical models, where $V_1^{Da_1}(q^2)$ is the vector form factor appearing in the $D \to a_1$ s-wave transition. The estimates are of the desired order of magnitude. We also predict branching fractions of its counterpart decays $D^0 \to K^{*-} a_1^+$ and $D^0 \to \bar{K}^{*0} a_1^0$.

## 1. Introduction

The $a_1(1260)$ resonance plays an important role in many phenomena of the nuclear and particle physics, but its basic properties are not very well known. The $a_1(1260)$ resonance mass and its decay width determined from different processes or by different experimental groups often contradict one another [1]. For instance, whereas the Particle Data Group [2] reports its decay width, $\Gamma_{a_1}$, to lie between 250 MeV and 600 MeV, a recent COMPASS measurement [3] provides a much smaller error $\Gamma_{a_1} = 367 \pm 9^{+28}_{-25}$ MeV. Due to the large decay width of the $a_1(1260)$ resonance, it can perform enigmatic tasks, thereby show discrepancy from naive expectations. It has been noticed for a long time that the predicted $D^0 \to K^- a_1^+$ and $D^+ \to \bar{K}^0 a_1^+$ rates are too small by a factor of 6 and 2, respectively, when compared with experiment [4]. Interestingly, branching fraction of a typical decay of charm meson $D^+$ emitting $a_1^+$ has been reported to be $B(D^+ \to \bar{K}^{*0} a_1^+) = (9.1 \pm 1.8) \times 10^{-3}$ in the Particle Data Properties [2]. The latter decay is supposed to be kinematically forbidden, if the peak values of particle masses are taken, since mass of the products (~2.15 GeV) is larger than the $D$ meson mass (1.86 GeV). However, due to the very large width of $a_1$ axial vector meson (250 - 600 MeV) and that of K* vector meson (~50 MeV), occurrence of these resonances has been seen in $D^+ \to \bar{K} \, 4\pi$ decay data [5]. We investigate this particular decay using factorization scheme in the Standard Model with a Breit-Wigner resonance formula. Earlier efforts have been made to study such smearing effects due to large $a_1$-width on $D \to \bar{K} a_1$ decays, which seem to improve the agreement with the experiment [6-9].

In this work, we focus on the W-emission diagrams to evaluate $D^+ \to \bar{K}^{*0} a_1^+$ decay amplitude as a function of the invariant mass of the $a_1$ meson. For this calculation, we treat the K* vector meson as a stable particle, as its decay width to mass ratio (< 6%) is much less than 30%-50% for the $a_1$ meson, and also to highlight the role of large decay width of $a_1$ resonance. The running decay width of $D^+ \to \bar{K}^{*0} a_1^+$ is then averaged over using the relativistic Breit-Wigner resonance formula. We have employed this procedure to predict the branching fraction of the similar decays of $D^0$ mesons. The uncertainty in the theoretical prediction follows mainly from the non-availability of the vector form factor in $D \to a_1$ transition, $V_1^{Da_1}(0)$, for which theoretical estimates differ by a factor of three or more [9-11].

## 2. Factorization Method

The relevant weak Hamiltonian needed to describe $D \to \bar{K} a_1$ decays in the Cabibbo-Kobayashi-Maskawa (CKM) favored mode is written as:

$$H_W = \frac{G_F}{\sqrt{2}} V_{ud} V_{cs}^* [a_1 (\bar{u}d)(\bar{s}c) + a_2 (\bar{s}d)(\bar{u}c)], \qquad (1)$$

where $(\bar{q}q)$ is a shorthand for the color singlet combination $\bar{q}\gamma_\mu (1-\gamma_5) q$ and $V_{ud} V_{cs}^*$ is the CKM factor. The coefficients, $a_1$ and $a_2$, which enclose QCD short distance effects at the charm quark mass scale, are taken to be $a_1 = 1.26$ and $a_2 = -0.51$ [8, 12].

For calculating the decay amplitudes, we use the following notations: $V \equiv 1^-$ vector meson, and $A \equiv 1^+$ axial-vector meson. The factorization scheme expresses the decay amplitudes as a product of the matrix elements of weak currents. Dropping the non-kinematic factors ($G_F / \sqrt{2}$, CKM, color, and flavor isospin C.G. factors), the amplitudes can be written as

$$A(D \to VA) \approx \langle V | V^\mu | 0 \rangle \langle A | J_\mu | D \rangle + \langle A | A^\mu | 0 \rangle \langle V | J_\mu | D \rangle. \qquad (2)$$

The explicit expressions for $V^\mu$, $A^\mu$, $J^\mu$ are understood. Using Lorentz invariance, the matrix elements of the current between meson states can be expressed as,

$$\langle V(k_V, \varepsilon^V) | V_\mu | 0 \rangle = \varepsilon_\mu^{V*} m_V f_V, \quad \langle A(k_A, \varepsilon^A) | A_\mu | 0 \rangle = \varepsilon_\mu^{A*} m_A f_A. \qquad (3)$$

Using the Bauer, Stech and Wirbel (BSW) model notation [12,13], the $D \to V$ transitions are given in terms of the following dimensionless form factors:

$$\langle V(k_V, \varepsilon^V) | V_\mu | D(p_D) \rangle = -\frac{1}{m_D + m_V} \varepsilon_{\mu\nu\alpha\beta} \varepsilon^{V*\nu} P^\alpha q^\beta V^{PV}(q^2),$$

$$\langle V(k_V, \varepsilon^V) | A_\mu | D(p_D) \rangle = +i \left\{ (m_D + m_V) \varepsilon_\mu^{V*} A_1^{PV}(q^2) - \frac{\varepsilon^{V*} \cdot P}{m_D + m_V} P_\mu A_2^{PV}(q^2), \right.$$

$$\left. - 2m_V \frac{\varepsilon^{V*} \cdot P}{q^2} q_\mu [A_3^{PV}(q^2) - A_0^{PV}(q^2)] \right\},$$

(4)

where $P_\mu = (p_D + k_V)_\mu$, $q_\mu = (p_D - k_V)_\mu$,

$$A_3^{PV}(q^2) = \frac{m_D + m_V}{2m_V} A_1^{PV}(q^2) - \frac{m_D - m_V}{2m_V} A_2^{PV}(q^2),$$

and

$$A_3^{PV}(0) = A_0^{PV}(0).$$

Similarly, the $D \to A$ transitions are expressed in terms of the dimensionless form factors [10,11] as

$$\langle A(k_A, \varepsilon^A) | A_\mu | D(p_D) \rangle = -\frac{1}{m_D + m_A} \varepsilon_{\mu\nu\alpha\beta} \varepsilon^{A*\nu} P^\alpha q^\beta A^{PA}(q^2), \tag{5}$$

$$\langle A(k_A, \varepsilon^A) | V_\mu | D(p_D) \rangle = +i\left\{ (m_D + m_A) \varepsilon_\mu^{A*} V_1^{PA}(q^2) - \frac{\varepsilon^{A*} \cdot P}{m_D + m_A} P_\mu V_2^{PA}(q^2) \right.$$

$$\left. - 2m_A \frac{\varepsilon^{A*} \cdot P}{q^2} q_\mu [V_3^{PA}(q^2) - V_0^{PA}(q^2)] \right\}, \tag{6}$$

where $P_\mu = (p_D + k_A)_\mu$, $q_\mu = (p_D - k_A)_\mu$,

$$V_3^{PA}(q^2) = \frac{m_D + m_A}{2m_A} V_1^{PA}(q^2) - \frac{m_D - m_A}{2m_A} V_2^{PA}(q^2)$$

and

$$V_3^{PA}(0) = V_0^{PA}(0).$$

The form factors appearing in (5) and (6) are not available in the BSW model, but have been calculated in the ISGW quark model [14] and its improved variant, ISGW2 model [15], which express the transition as,

$$\langle A(k_A, \varepsilon^A) | J_\mu | D(p) \rangle = l\varepsilon_\mu^{A*} + c_+(\varepsilon^{A*} \cdot p_D)(p_D + k_A)_\mu + c_-(\varepsilon^{A*} \cdot p_D)(p_D - k_A)_\mu$$

$$-q\varepsilon_{\mu\nu\alpha\beta}\varepsilon^{A*\nu}(p_D+k_A)^{\alpha}(p_D-k_A)^{\beta}, \tag{7}$$

where $l$, $c_{+,-}$ and $q$ are another set of the form factors. It is straightforward to see that the two sets of form factors are related as:

$$A^{PA}=-(m_D+m_A)q,\ V_1^{PA}=-il/(m_D+m_A),\ V_2^{PA}=i(m_D+m_A)c_+,$$

and 
$$V_0^{PA}=\frac{-i}{2m_A}[l+(m_D^2-m_A^2)c_++q^2c_-].$$

Note that in $D\to VA$ decays, similar to $D\to VV$ decays, in general $s-$, $p-$ and $d-$waves can contribute to the relative angular momentum between the final states. Leaving aside the non-kinematic factors, the decay amplitudes thus become [16-18],

$$A(D\to VA)=m_A f_A\Big\{i(m_D+m_V)\varepsilon^{V*}\cdot\varepsilon^{A*}A_1^{PV}(q^2)+\frac{2}{m_D+m_V}\varepsilon_{\mu\nu\alpha\beta}\varepsilon^{V*\mu}\varepsilon^{A*\nu}p_D^{\alpha}k_V^{\beta}V^{PV}(q^2)$$

$$-i\frac{1}{m_D+m_V}\varepsilon^{V*}\cdot(p_D+k_V)\varepsilon^{A*}\cdot(p_D+k_V)A_2^{PV}(q^2)\Big\}+\{V\leftrightarrow A\}, \tag{8}$$

where the 1st, 2nd, and 3rd terms correspond to $s-$, $p-$ and $d-$waves respectively. The decay width formula is then expressed as,

$$\Gamma(D\to VA)=(nonkinematic-factors)\frac{k}{8\pi m_D^2}\Big|\sum_{i=s,p,d}decay-amplitude(i)\Big|^2. \tag{9}$$

However, it is naturally expected that the decay amplitudes for $p-$wave and $d-$wave are highly suppressed due to the lack of phase space in $D\to K^*a_1$ decays. Indeed, it has been shown earlier that even for a kinematically allowed $D\to VV$ process, the contributions from $p-$wave and $d-$wave are suppressed in comparison to $s-$wave by kinematic factors of around 75 [16-18]. Moreover, for heavier $B$ meson decays, the CLEO Collaboration [19,20] has performed the first full angular analysis of $B\to J/\psi K^{*0}$ and $J/\psi K^{*+}$ decays, which has confirmed that the $p-$wave component is significantly small in these decays. Thus the retention of $s-$wave only appears to be a reasonable approximation for the decays under consideration. The decay rate is then simplified as

$$\Gamma(D\to VA)=\frac{k}{8\pi m_D^2}\left[2+\left(\frac{m_D^2-m_V^2-m_A^2}{2m_V m_A}\right)\right]|A(D\to VA)|^2, \tag{10}$$

besides the non-kinematic factors, where

$$A(D \to VA) = i \{m_V f_V (m_D + m_A) V_1^{PA}(q^2) + m_A f_A (m_D + m_V) A_1^{PV}(q^2)\} \quad (11)$$

and $k$, the magnitude of the three-momentum of the final states, is given by

$$k = \frac{\sqrt{(m_D^2 - (m_V + m_A)^2)(m_D^2 - (m_V - m_A)^2)}}{2m_D}. \quad (12)$$

For the $A_1^{PV}$ form factor, BSW model [12,13] gives $A_1^{D\bar{K}^*}(0) = 0.88$, whereas its value $0.61 \pm 0.05$ has been obtained from the semileptonic decays $D \to \bar{K}^* + l + \nu_l$ [21]. Therefore, we take the averaged value $0.74$. Since $q^2-$ dependence of the form factors is difficult to extract either from theory or experiment, generally, the nearest pole dominance is assumed [12],

$$A_1^{D\bar{K}^*}(q^2) = \frac{A_1^{D\bar{K}^*}(0)}{(1 - q^2/m_A^2)^n}, \quad (13)$$

with $m_{A(1^+)} = 2.53 GeV$ being the pole mass corresponding to the weak current $(\bar{s}c)$ responsible for the $D \to \bar{K}^*$ transitions. The BSW model [12,13] assumes a monopole behavior ($n=1$), which we also adopt for obtaining conservative estimate. However, it has been pointed [9] that consistency with the heavy quark symmetry demands certain form factors to have dipole $q^2-$ dependence in the modified BSW model. We observe that choosing dipole behavior ($n=2$) would increase $B(D^+ \to \bar{K}^{*0} a_1^+)$ by 30%.

Since BSW provides the form factors only for $P(0^-) \to P(0^-)$ and $P(0^-) \to V(1^-)$ transitions, we take other form factors $V_1^{DA}$ from the improved ISGW2 quark model [15]. Following the analysis of Ref. [9], we take $V_1^{Da_1}(m_K^2) = -0.42$. Assuming the $q^2$ dependence similar to (13), we obtain $V_1^{Da_1}(0) = -0.40$, with corresponding pole mass $m_{V(1^-)} = 2.01 GeV$ for $(\bar{d}c)$ weak current.

For the decay constants, we choose $f_{a_1} \approx f_{K^*} = 0.221\ GeV$, though an experimental value of $f_{a_1} = 0.203 \pm 0.018\ GeV$, is quoted in [22]. It is generally argued that $a_1(1260)$ should have a similar decay constant as the $\rho-$ meson.

Sandwiching the weak Hamiltonian (1) between the initial and the final states, the decay amplitudes for $D \to \bar{K}^* a_1$ are obtained. The amplitude can be classified into three categories:

1. color favored amplitude involving the $D \to \overline{K}^*$ form factor,

$$A(D^0 \to K^{*-}a_1^+) = \frac{G_F}{\sqrt{2}} V_{ud} V_{cs}^* \, i\, a_1 \, m_{a_1} \, f_{a_1}(m_D + m_{K^*}) A_1^{DK^*}(m_{a_1}^2);$$

(14a)

2. color suppressed amplitude involving the $D \to a_1$ form factor,

$$A(D^0 \to \overline{K}^{*0}a_1^0) = \frac{G_F}{\sqrt{2}} V_{ud} V_{cs}^* \frac{i}{\sqrt{2}} a_2 \, m_{K^*} \, f_{K^*}(m_D + m_{a_1}) V_1^{Da_1}(m_{K^*}^2);$$

(14b)

3. amplitudes involving the both $D \to \overline{K}^*$ and $D \to a_1$ form factors,

$$A(D^0 \to K^{*-}a_1^+) = \frac{G_F}{\sqrt{2}} V_{ud} V_{cs}^* \, i\{a_1 \, m_{a_1} \, f_{a_1}(m_D + m_{K^*}) A_1^{DK^*}(m_{a_1}^2)$$

$$+ a_2 \, m_{K^*} \, f_{K^*}(m_D + m_{a_1}) V_1^{Da_1}(m_{K^*}^2)\}.$$

(14c)

The possible W-annihilation diagram is neglected due to color and helicity suppression. Within the approximation we take, the first two amplitudes consist only of a single term, while the third one includes two terms. Thus the branching ratio of the third one is affected by the signs of the form factors and decay constants.

## 3. Smearing Effects due to Wide Width of $a_1$ Resonance

Obviously treating $a_1$ as a sharp resonance will not permit the decays due to the lack of the phase space. Therefore, we take into account the so-called smearing effects caused by the large $a_1$ width. Since the $a_1$ width is rather wide, the phase space effectively becomes available for the final state in $D \to \overline{K}^* a_1$ decays. Let us denote the center-of-mass energy of $a_1$ in its rest frame as $m$, and take the average for the decay rate over $m$ using a measure, i.e.,

$$\int \rho(m) \Gamma(D \to \overline{K}^* a_1(m)) dm.$$

(15)

We employ a relativistic "Breit-Wigner" measure

$$\rho(m) = \frac{2m}{\pi} \frac{m \Gamma_{tot}(m)}{(m^2 - m_{a_1}^2) + m^2 \Gamma_{tot}^2(m)},$$

(16)

where $m_{a_1}$ is the experimental value of $a_1$ mass, $1.230 GeV$. Here, the total width of $a_1$ is parameterized with respect to its dominant decay mode $a_1 \to \rho\pi$ to get

$$\Gamma_{tot}(m) = \alpha(m) \Gamma_{tot} \theta(m - m_\rho - m_\pi).$$

(17)

$\Gamma_{tot}$ is the total width of $a_1$, which is taken to vary from 0.300 GeV to 0.600 GeV following the

Particle Data Group [2]. The step function ensures that the $\rho\pi$ channel opens only when the $a_1$ mass is above the threshold. $\alpha(m)$ is the kinematic factor defined by

$$\alpha(m) = \frac{k(m)}{k(m_{a_1})}, \tag{18}$$

where $k$, the momentum of $\rho$ and $\pi$ in the $a_1$ rest frame, is given by.

$$k(m) = \frac{\sqrt{(m^2 - (m_\rho + m_\pi)^2)(m^2 - (m_\rho - m_\pi)^2)}}{2m}. \tag{19}$$

For the sake of simplicity and maintaining clarity, we have ignored the possible but small effect from the $\overline{K}^*$ width, as it is 6% or less of its mass in comparison to 30-50% for $a_1(1260)$ resonance. In obtaining the decay amplitudes, we use the invariant mass $m$ in the form factor $A_1^{D\overline{K}^*}(m^2)$ defined in Eq. (13) and follow ref. [6] to scale the decay constant $f_{a_1}(m)$ as,

$$f_{a_1}(m) = f_{a_1} \left(\frac{m_{a_1}}{m}\right)^{1/2}. \tag{20}$$

## 4. Numerical Results and Discussion

With the method and the numerical values given above, branching ratio of the charged $D^+$ meson is estimated to be $B(D^+ \to \overline{K}^{*0} a_1^+) = 3.3 \times 10^{-3}$ for $\Gamma_{tot}$ = 0.4 GeV, which is clearly smaller than the experimental value $B_{exp}(D^+ \to \overline{K}^{*0} a_1^+) = 9.1 \pm 1.8 \times 10^{-3}$. Here we should emphasize that the purpose of this first trial is not to reproduce the precise value of the branching ratio, but rather to see what comes out from a naïve estimate for future reference. As a consequence, it is found that a naïve estimate provides the right order of magnitude. In the rest of this section, we show the estimate with the parameters varying in reasonable ranges to explore the potential size of uncertainty.

We first examine the $\Gamma_{tot}$ dependence of the branching ratios by fixing the unknown $D \to a_1$ form factor to be $V_1^{Da_1}(0) = -0.40$. The numerical results of $B(D^+ \to \overline{K}^{*0} a_1^+)$ for different values of $\Gamma_{tot}$ = 0.3, 0.4, 0.5, 0.6 GeV are listed in Tab. 1, which shows an increase almost linear in $\Gamma_{tot}$ up to $B(D^+ \to \overline{K}^{*0} a_1^+) = 4.8 \times 10^{-3}$.

We took the negative value for one of the form factor, $V_1^{Da_1}(0) = -0.40$. Remembering $a_2$ is negative, the estimates in Tab. 1 are said to be the results for constructive interference. Unfortunately, in the factorization approximation, the nature of the interference is totally out of control, and destructive interference is also possible. Indeed, for other two-body decays such as $D^+ \to \overline{K}^0\pi^+ / \overline{K}^{*0}\pi^+ / \overline{K}^0\rho^+ / \overline{K}^{*0}\rho^+$, destructive interference has been observed [13, 23]. Thus, we calculated the $B(D^+ \to \overline{K}^{*0}a_1^+)$ in the presence of destructive interference, and found that it is reduced by half. Whether constructive or destructive interference, theoretical value, in general, is less than the experimental one, indicating the importance of nonfactorizable effects to this decay mode like in other charm meson decays [24-26].

Among various possible sources of uncertainties, the uncertainty of the form factor $V_1^{Da_1}(0)$ presumably dominates the others besides the one from the factorization approximation. Indeed, in some models [10, 11], the form factor $V_1^{Da_1}(0)$ turns out to be positive and large, 1.50. Since it is positive, the nature of the interference appears to be changed. However, in those models, sign of $a_1$ decay constant is negative, and hence again the constructive interference is realized in this case. For $\Gamma_{tot} = 0.4 GeV$, and $V_1^{Da_1}(0) = 1.50$, we obtain $B(D^+ \to \overline{K}^{*0}a_1^+) = 7.0 \times 10^{-3}$, which is quite close to the experimental value.

We have also calculated branching fractions of such decay modes of $D^0$ meson, which are tabulated in Tab. 1 for $V_1^{Da_1}(0) = -0.40$ and $1.50$. Obviously, for $V_1^{Da_1}(0) = 1.50$, the branching fraction of $D^0 \to \overline{K}^{*0}a_1^0$ gets enhanced by a factor of 14, say giving $B(D^0 \to \overline{K}^{*0}a_1^0) = 0.21 \times 10^{-3}$ for $\Gamma_{tot} = 0.4 GeV$, whereas $B(D^0 \to K^{*-}a_1^+)$ remains unaffected. Again these first estimates can be used as a reference value when more realistic or sophisticated analysis is available in future.

**Acknowledgments:**
RCV thanks for the hospitality and support at the KEK Theory Center, High Energy Accelerator Research Organization (KEK). Financial assistance from the CSIR (New Delhi) under Professor Emeritus Scheme awarded to him is also gratefully acknowledged.

**References**

1. M. Vojík and P. Lichard, "Three-pion decays of the tau lepton, the $a_1(1260)$ properties, and the $a_1\rho(1260)\pi$ Lagrangian", [arXiv:1006.2919v1][hep-ph].
2. K.A. Olive et al. (Particle Data Group), Chin. Phys. C, **38**, 090001 (2014).
3. COMPASS collaboration, M. Alekseev et al., Phys. Rev. Lett. 104, (2010) 241803. [arXiv:0910.5842] [INSPIRE].
4. D.M. Coffman et al., Phys. Rev. D **45**, 2196 (1992).
5. FOCUS Collaboration, J.M. Link et al., Phys. Lett. B **561**, 225 (2003).
6. A.N. Kamal and R.C. Verma, Phys. Rev. D **45**, 982 (1992).
7. T.N. Pham, Phys. Rev. D **46**, 2976 (1992).
8. A.C. Katoch and R.C. Verma, J. Phys. G **21**, 525 (1995).
9. H.Y. Cheng, Phys. Rev D **67**, 094007 (2003).
10. H.Y. Cheng, C.K. Chua, and C.W. Hwang, Phys. Rev. D **69**, 074025 (2004).
11. R.C. Verma, J. Phys. G **39,** 025005 (2012).
12. M. Wirbel, Prog. Part. Nucl. Phys. **21**, 33 (1988).
13. M. Wirbel, B. Stech, and M. Bauer, Z. Phys. C **29**, 637 (1985).
14. N. Isgur, D. Scora, B. Grinstein, and M.B. Wise, Phys. Rev. D **39**, 799 (1989).
15. D. Scora and N. Isgur, Phys. Rev. D **52**, 2783 (1995).
16. A.N. Kamal, R.C. Verma, and N. Sinha, Phys. Rev. D **43**, 843 (1991).
17. El Hassan, El Aaoud, and A. N. Kamal, Phys. Rev. D **59**, 114013 (1999).
18. R. Mohanta, A.K. Giri, M. P. Khanna, M. Ishida, S. Ishida and M. Oda, Prog. Theor. Phys. **101**, 947 (1999).
19. CLEO Collaboration, C. P. Jessop et al., Phys. Rev. Lett. **79**, 4533 (1997).
20. X. G. He and W. S. Hou, Phys. Lett. B **445**, 344 (1999).
21. M.S. Witherell, `Charm Decay Physics' in "Lepton and Photon Interaction", Proc. of 16th Int. Sym. Ithaca, NY 1993, ed. P. Drell and D. Rubin, AIP Conf. Proc. No. 302, (AIP, NY, 1994).
22. J.C.R. Bloch, Yu.L. Kalinovsky, C.D. Roberts, and S.M. Schmidt, Phys. Rev. D **60**, 111502 (1999).
23. T.E. Browder, K. Honscheid and D. Pedrini, Annu. Rev. Nucl. Part. Sci. **46**, 395 (1996).
24. H.Y. Cheng, Phys. Lett. B **335,** 428 (1994),
25. A.N. Kamal, A.B. Santra, T. Uppal, and R.C. Verma, Phys. Rev. D **53,** 2506 (1996)
26. R. Dhir and R.C.Verma, J. Phys. G. **34,** 637 (2007).

**Table 1:** Branching fractions of $D \to \bar{K}^* a_1$ decays ($\times 10^{-3}$) for different values of $a_1$ total width ($\Gamma_{tot}$) and $V_1^{Da_1}(0) = -0.40$. Values given in the parentheses are for $V_1^{Da_1}(0) = 1.50$.

| | $\Gamma_{tot} = 0.3 GeV$ | $\Gamma_{tot} = 0.4 GeV$ | $\Gamma_{tot} = 0.5 GeV$ | $\Gamma_{tot} = 0.6 GeV$ | Exp. |
|---|---|---|---|---|---|
| $D^+ \to \bar{K}^{*0} a_1^+$ | 2.5 (5.3) | 3.3 (7.0) | 4.1 (8.7) | 4.8 (10.2) | 9.1±1.8 |
| $D^0 \to K^{*-} a_1^+$ | 0.6 (0.6) | 0.8 (0.8) | 1.0 (1.0) | 1.2 (1.2) | --- |
| $D^0 \to \bar{K}^{*0} a_1^0$ | 0.011 (0.16) | 0.015 (0.21) | 0.019 (0.26) | 0.022 (0.31) | --- |